\documentstyle[aps,prc,preprint,epsf,twoside]{revtex}

\newcommand{\sdj}[6]{ \left( \begin{array}{*{3}{c}}
                                      #1 & #2 & #3 \\
                                      #4 & #5 & #6 
                                    \end{array} \right) }

\newcommand{\bra}[1]{\left\langle{#1}\right|}
\newcommand{\ket}[1]{\left|{#1}\right\rangle}
\newcommand{\braket}[2]{\left\langle{#1}\right.\left|{#2}\right\rangle}

\newcommand{\bramket}[3]{\left\langle\,{#1}\,\left|\,{#2}\,
            \right|\,{#3}\,\right\rangle}

\newcommand{\intp}[1]{\int\frac{d^{3}{#1}}{(2\pi)^{3}}}

\newcommand{\bqn}{\begin{eqnarray}}
\newcommand{\eqn}{\end{eqnarray}}
\newcommand{\Bqnn}{\begin{eqnarray*}}
\newcommand{\Eqnn}{\end{eqnarray*}}
\newcommand{\nn}{\nonumber\\}

\newcommand{\Bqn}{\begin{eqnarray}}
\newcommand{\Eqn}{\end{eqnarray}}
\newcommand{\beq}{\begin{equation}}
\newcommand{\eeq}{\end{equation}}
\newcommand{\ov}{\overline}
\newcommand{\eps}{\epsilon}

\newcommand{\half}{\frac{1}{2}}

\begin{document}
\title{A Spectator-Quark-Model for the Photoproduction of Kaons}
\author{V. Keiner}
\address{Institut f\"ur Theoretische Kernphysik,\\
         Universit\"at Bonn, Nussallee 14-16, 53115 Bonn, FRG}
\date{January 19, 1995}
\maketitle

\begin{abstract}

A simple model for the photoproduction of kaons off protons
with a lambda hyperon in the final state is presented. In a
quark model, the interaction is modelled by the pair-creation 
of the (anti-) strange quarks in the final state which recombine with
the three quarks of the proton to form the lambda and kaon. 
The calculated scattering cross sections for photon energies up to
$E_\gamma = 1.9 \; \mbox{GeV}$ are compared to experiment. 
The pair-creation process is found to have a significant
contribution to the total cross section. 

\end{abstract}

\narrowtext
\newpage

\section{Introduction}

\begin{figure}
\begin{center}
\leavevmode
\epsfxsize=0.7\textwidth
\epsffile{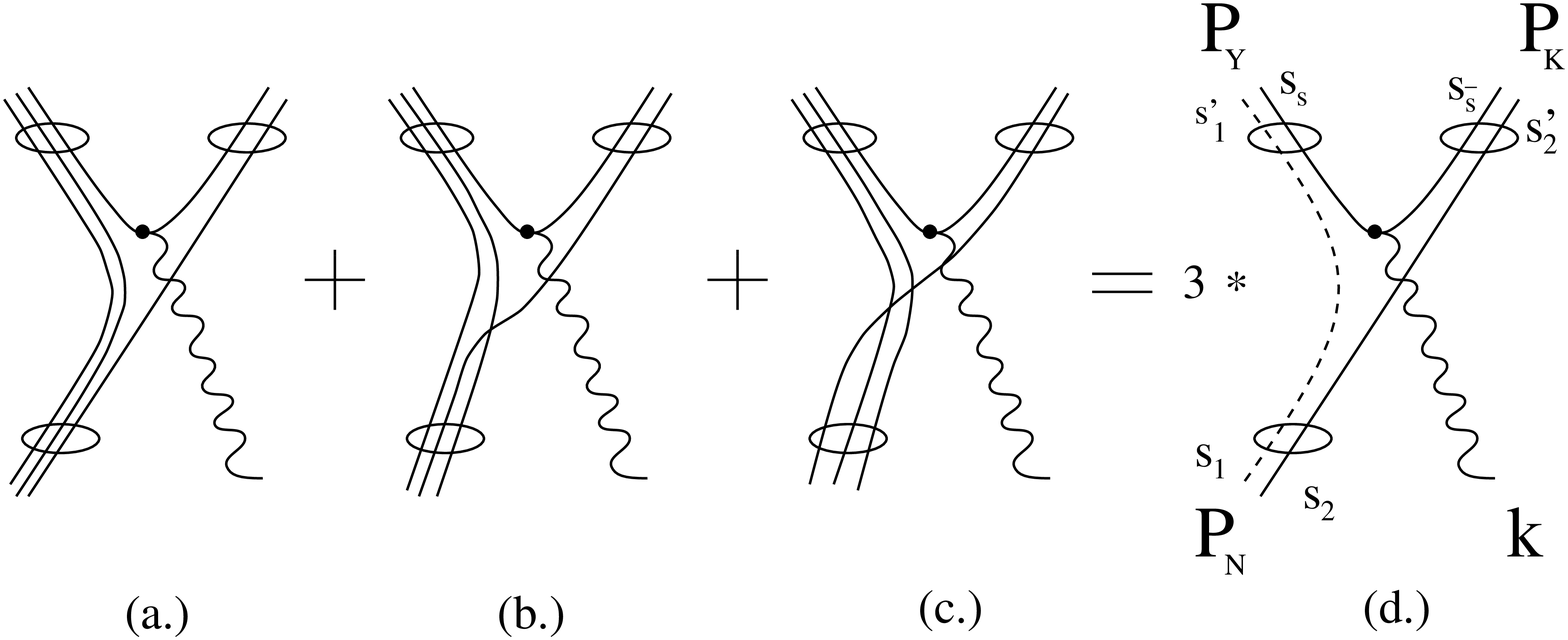}
\end{center}
\caption{Photoproduction of mesons by quark-antiquark pair-creation; 
  the photon couples to the 
  \mbox{(anti-)}strange quarks, the u-d-quark pair  
  and the u-quark are spectator particles; (d.) is the only diagram
  after correctly symmetrizing the baryonic wave functions.}
\label{proc}
\end{figure}

In the last years, there has been renewed interest in the
photoproduction of strange particles off protons. This is 
especially due to powerful new facilities (SAPHIR at ELSA (Bonn) 
\cite{schwille} , CLAS at CEBAF in the near future) 
which provide new data with better statistics. 
The process $\gamma + p \rightarrow K^+ + \Lambda/\Sigma^0$
is a useful tool to study strangeness and its production
in hadronic matter. From the measured cross sections one can
extract strong coupling constants and magnetic moments of 
the produced hyperons. Of special interest are polarization 
observables to study the spin-dependences of the reaction.
Many efforts have 
been made to describe the process 
$\gamma + p \rightarrow K^+ + \Lambda/\Sigma^0$ for medium 
energies (up to photon energies $E_\gamma=2 \; \mbox{GeV}$). Good results
for the cross sections have been achieved with various isobaric
models, of which the work of Adelseck et al. \cite{adelseck} is the
latest example.  
However, Adelseck et al. have been unable to explain
the polarization data, like the recoil polarization of the
hyperon. In addition, their calculated total cross section
rises for photon energies larger than 1.4 \mbox{GeV}, in contrast
to experiment \cite{schwille,schoch}.
One might expect that the observed recoil polarization
is a direct consequence of the quark structure of hadronic matter,
as has been proposed by Miettinen \cite{miett} for the strangeness
production in pp collisions. In the naive quark model, where
a baryon is composed of three quarks, the spin of the $\Lambda$ 
is carried by the strange quark alone, in contrast to the 
$\Sigma^0$, where u and d quark couple to a spin triplet. 
Thus, it is a challenging task to explain the recoil polarization 
in the framework of a simple quark model. \\
This paper describes a simple model of the process
$\gamma + p \rightarrow K^+ + \Lambda$
for energies up to $E_\gamma=1.9 \; \mbox{GeV}$. The baryons are composed
of three quarks ($\Lambda=(uds)$, $p=(uud)$) and the $K^+$ meson of
a quark-antiquark pair ($u \ov s$). The only contribution being considered
(see fig.\ref{proc}) describes the reaction by a pair-creation of
the strange quark and antiquark which recombine with the up quark to the
$K^+$ and the up and down quarks to the $\Lambda$, respectively. 
The pair-creation process, however, does not describe the $\Sigma^0$
production. This can be seen by considering forward
scattering, where obviously only the $M_{S_{12}}=0$ component
of the $S_{12}=1$ (ud) state contributes. Thus, the $\Sigma^0$
production amplitude is suppressed by a factor 1/3.
The baryon states are described by an integral over
quark states and a Gaussian function and the meson state by
that over quark-antiquark states and a Gaussian function 
\cite{hayne,vanroyen}. 

\section{The Model}

The antisymmetrized baryon wave function has the following structure, 
e.g. for 
the proton
\bqn
\ket{N_{s}(P_N)} & = & N_\alpha \sqrt{2 P_N^0} \; 
\intp{p_\rho} \, \intp{p_\lambda} \, 
 \frac{1}{\sqrt{2p_1^0 \, 2p_2^0 \, 2p_3^0}} \, 
 {\cal R}_\alpha(p_\rho, p_\lambda) \; \chi_N^F \chi_N^C \, \nn
 & & \hspace{3cm} \cdot
     \left[ \left[ \ket{\vec p_1, s_1} \otimes \ket{\vec p_2, s_2}
     \right]^{s_{12}} \otimes \ket{\vec p_3, s_3} \right]^s
\eqn
with
\bqn
\vec p_\rho & = & \frac{m_2}{m_1+m_2} \; \vec p_1 - 
                  \frac{m_1}{m_1+m_2} \; \vec p_2 \nn
\vec p_\lambda & = & \frac{1}{m_1+m_2+m_3} \;
( m_3 \; (\vec p_1 + \vec p_2)-(m_1+m_2) \vec p_3 ) \quad.
\eqn
The various functions and parameters will be explained below.
After symmetrizing the wave function with respect to the 
interchange of any two quarks in the baryon, it suffices to 
calculate only graph (a) in fig.\ref{proc} and multiply by 3, which gives
diagram (d.) (see appendix D).
The graph invokes the idea that the quark pair may form a
diquark. Indeed, after correctly symmetrizing, one 
can integrate in the $\cal T$ matrix, see below, over $\vec p_\rho$,
thus replacing $\ket{\vec p_1} \, \ket{\vec p_2}$ by $\ket{\vec p_{12}}$.
In our model, the internal dynamics of the (ud) pair does not
affect the scattering process. From now on, the index 1 denotes the
spectator (ud) pair.
We thus can write for the hadron wave functions (note that for
$\Lambda$ production the diquark has spin 0): \\
The proton wave function is
\bqn
\ket{N_{s_2}(P_N)} = N_\alpha \sqrt{2 P_N^0} \; \intp{p} \, 
 \frac{1}{\sqrt{2p_1^0 \,  2p_2^0}} \, 
 {\cal R}_\alpha(p) \; \chi_N^F \chi_N^C \; \ket{\vec p_1} \ket{\vec
   p_2, s_2}
\eqn
with the (di-)quark states
\bqn 
\ket{\vec p_1} & = & \ket{\frac{m_1}{m_1+m_n} \vec P_N+\vec p} 
                =  \tilde a^{\dag} (p_1) \; \ket{0} \nn 
\ket{\vec p_2, s_2} & = & \ket{\frac{m_n}{m_1+m_n} 
                                \vec P_N-\vec p, s_2} 
                =  a_{s_2}^{\dag} (p_2) \; \ket{0} \quad, 
\eqn  
the Gaussian function
\bqn
{\cal R}_\alpha(p) = e^{-\alpha^2 p^2} \quad,\quad  p = |\vec p\,| \nonumber
\eqn 
and $\chi^F$, $\chi^C$ the flavour and colour functions with the
correct symmetry for the ground state. 
Analogously, we have for the $\Lambda$ wave function
\bqn
\bra{Y_{s_s} (P_Y)} = N_\beta \, \sqrt{2 P_Y^0} \, \intp{p'} \, 
 \frac{1}{\sqrt{2{p'}_1^0 \,  2p_s^0}} \,
 {\cal R}_\beta(p') \; \chi_Y^F \chi_Y^C \; \bra{\vec p_1\,'}
  \; \bra{\vec p_s, s_s} 
\eqn
with
\bqn
\bra{\vec p_1\,'} & = & \bra{\frac{m_1}{m_1+m_s} \vec P_Y+\vec p\,'} 
                = \bra{0}  \, \tilde a (p'_1) \nn
\bra{\vec p_s, s_s} & = & \bra{\frac{m_s}{m_1+m_s} 
                                \vec P_Y-\vec p\,', s_s} 
                =  \bra{0} \, a_{s_s} (p_s)  
\eqn  
and the $K$ meson state is written as
\bqn
\bra{K (P_K)} = \sum_{s_{\ov s} s_2'} \, C_{\half s_{\ov s} \half
  s_2'}^{0 0} \,  
 N_\gamma \, \sqrt{2 P_K^0} \, \intp{p''} \, 
 \frac{1}{\sqrt{2p_{\ov s}^0 \,  2{p'}_2^0}} \,
 {\cal R}_\gamma(p'') \; \chi_K^F \chi_K^C \; 
 \bra{\vec p_{\ov s}, s_{\ov s}} \; \bra{\vec p_2\,', s_2'}
\eqn
with
\bqn
\bra{\vec p_{\ov s}, s_{\ov s}} & = & \bra{\frac{m_s}{m_s+m_n} 
                                \vec P_K+\vec p\,'', s_{\ov s}} 
                =  \bra{0} \, b_{s_{\ov s}} (p_{\ov s}) \nn 
\bra{\vec p_2\,', s_2'} & = & \bra{\frac{m_n}{m_s+m_n} 
                  \vec P_K-\vec p\,'', s_2'}  
                = \bra{0}  \, a_{s_2'} (p'_2) \quad.
\eqn  
The decomposition of the field operators $\Psi, \ov\Psi$ 
is given in appendix B. The normalization reads
\bqn
<\vec k, s|\vec k\,', s'> \; = \; (2 \pi)^3 \, (2 k^0) \, \delta^3(\vec k -
\vec k\,') \, \delta_{s s'}  \quad.
\eqn  
This leads to the standard normalization of the hadronic states
(see appendix A).

There are at least five parameters: the three oscillator parameters
of the proton, the $\Lambda$ and the $K^+$, respectively, and the constituent 
quark masses of the strange (s)  and non-strange (n) quarks. One may
introduce an effective mass of the spectator diquark being less
than twice the mass of the n-quark, which is favoured by
many diquark models \cite{lichtenberg}. In fact, the (ud) pair 
forming a scalar diquark (better: diquark correlation) 
is postulated by many authors, see \cite{anselmino} for a
review on that subject. However, we will see that the variation of
those parameters in a reasonable range does not
greatly affect the prediction of the model.

The following {$\cal T$} matrix is calculated. For $\Lambda$ 
production the spin of the spectator diquark is 
$s_1 = s'_1 = 0$, and thus the z-components of the proton and
$\Lambda$ spin are $s = s_2$ and $s' = s_s$, respectively.
\bqn
{\cal T} = \bramket{Y_{s_s}(P_Y), K(P_K)}
        {\ov \Psi(0) \, ie \, \hat Q \, \gamma^\mu \, \Psi(0) \, A_\mu(0)} 
        {N_{s_2}(P_N), (\vec k, \lambda)} \quad.
\eqn   
With the commutator relations of the creation and annihilation
operators (see appendix B) we find for real photons 
($\lambda = 1, \, 2$) (without colour and flavour factors):
\bqn
{\cal T} & = & N_\alpha N_\beta N_\gamma \, \sqrt{2P_N^0}
\sqrt{2P_Y^0} \sqrt{2P_K^0} \nn
 & & \intp{p} \, \frac{1}{\sqrt{2 p_s^0 \, 2 p_{\ov s}^0}} \, 
     {\cal R}_\alpha(p) {\cal R}_\beta(p'(p)) {\cal R}_\gamma(p''(p))
     \nn
 & & \sum_{s_{\ov s}} \, C_{\half s_{\ov s} \half s_2}^{0 0} \,
      \left( \ov u_{s_s}(p_s) \, ie \hat Q \, \gamma^k \, v_{s_{\ov
            s}}(p_{\ov s}) \, \eps_k^{(\lambda)} \right) \quad.
\eqn  
For the operator one gets 
\bqn
{\cal O} & := & \sum_{s_{\ov s}} \, C_{\half s_{\ov s} \half s_2}^{0 0} \;
               \ov u_{s_s}(p_s) \, \gamma^k \, v_{s_{\ov s}}(p_{\ov s}) \,
               \eps_k^{(\lambda)} \\
         & = & \sum_{s_{\ov s}} \, C_{\half s_{\ov s} \half s_2}^{0 0} \;
           (-1) \, \sqrt{p_s^0+m_s} \, \sqrt{p_{\ov{s}}^0+m_s} \;
           \chi_{s_s}^{\dag}  \left( \sigma^k + 
           \frac{\vec\sigma \vec p_s}{p_s^0+m_s} \, \sigma^k \,     
           \frac{\vec\sigma \vec p_{\ov s}}{p_{\ov s}^0+m_s} \right)
           \tilde \chi_{s_{\ov s}} \; \eps_k^{(\lambda)} 
           \quad. \nonumber 
\eqn
In the center-of-mass system, this can be written as
\bqn
{\cal O} 
& = & \sum_{s_{\ov s}} \, C_{\half s_{\ov s} \half s_2}^{0 0}
      \; \sum_{i=1}^{19} \, 4 \pi \, \tilde c_i \, p^{t_i} \, P_K^{v_i}
      \, k^{u_i} \nn
&   & \chi_{s_s}^{\dag} \,
      \left[ Y_{\tilde l_{2_i}}(\hat P_K) \otimes  
      \left[ \left[ Y_{l_{2_i}}(\hat p) \otimes
      \left[ \sigma^{a_i} \otimes Y_{l_{1_i}}(\hat p)
      \right]^{k_i} 
      \right]^{a_i} \otimes
      Y_{\tilde{l_{1_i}}}(\hat k)
      \right]^{\tilde k_i}
      \right]_k^1 \; \tilde \chi_{s_{\ov s}} \; \eps_k^{(\lambda)} 
\eqn      
with momentum-dependent coefficients $\tilde c_i = 
\tilde c_i(p_s^0,p_{\ov s}^0)$ . \\
Now define
\bqn 
\tilde m_1 & = & \frac{m_1}{m_1+m_n} \quad,\quad \tilde m_n =
\frac{m_n}{m_1+m_n} \nn 
\tilde{\tilde m}_1 & = & \frac{m_1}{m_1+m_s} \quad,\quad 
\tilde{\tilde m}_n =
\frac{m_n}{m_s+m_n} \nn
\delta & = & \sqrt{\alpha^2+\beta^2+\gamma^2} \nn
\alpha_1 & = & \frac{1}{\delta} \, (-\beta^2 \tilde m_1 + \gamma^2
\tilde m_n) \quad,\quad 
\alpha_2 = \frac{1}{\delta} \, (\beta^2 \tilde{\tilde m}_1 + \gamma^2
\tilde{\tilde m}_n)   \quad.                
\eqn
A decomposition with Clebsch-Gordan coefficients, integration
over $d\Omega_p$ (where we neglect the angular dependence of
$p_s^0$ and $p_{\ov s}^0$), identification 
of the z-direction with $\hat P_K$ and partial summation gives,
now for the general case of a spectator-diquark spin 
$S_1 = 0 \mbox{ or } 1$
\bqn
{\cal T} & = & \bramket{Y_{s'}(P_Y), K(P_K)}
        {\ov \Psi(0) \, ie \, \hat Q \, \gamma^\mu \, \Psi(0) \, A_\mu(0)} 
        {N_{s}(P_N), (\vec k, \lambda)} \nn
 & = & \sqrt{2P_N^0} \sqrt{2P_Y^0} \sqrt{2P_K^0} \, N_\alpha N_\beta
       N_\gamma \nn
 & &   \cdot \exp\left(-k^2(\beta^2 \tilde m_1^2+\gamma^2 \tilde
       m_n^2-\alpha_1^2) \right) \nn
 & &   \cdot \exp \left( -P_K^2 (\beta^2 \tilde{\tilde m}_1^2 + \gamma^2
       \tilde{\tilde m}_n^2 - \alpha_2^2 ) \right) \nn
 & &   \cdot \exp \left(-k \, P_K \, \cos \, \theta_K \; 
       (-2 \beta^2 \tilde m_1
       \tilde{\tilde m}_1 + 2 \gamma^2 \tilde m_n \tilde{\tilde m}_n
       -2 \alpha_1 \alpha_2 ) \right) \nn  
 & &   \cdot \sum_{i=1}^{19} \; {\cal I}_i  \; c_i \; k^{u_i} \; P_K^{v_i} \nn
 & &   \cdot \sum_{s_1 s_2 s_s m_s \tilde m_k \tilde m_{1_i} \tilde m_{2_i}}
       \sqrt{\frac{2 \tilde l_{2_i} +1}{4 \pi}} \, 
       \delta_{\tilde m_{2_i} 0} \nn
 & &   \hspace{2.5cm} \sqrt{\frac{2 \tilde l_{1_i} +1}{4 \pi}} \,
       \sqrt{\frac{(\tilde l_{1_i} - \tilde m_{1_i})!}{(\tilde l_{1_i}
           + \tilde m_{1_i})!}} \; P_{\tilde l_{1_i}}^{\tilde m_{1_i}}
           (\cos \, \theta_K) \nn
 & &   \cdot {\cal O}_{\tilde m_{1_i} \tilde m_{2_i} m_s \tilde m_k}^{i; s_1
   s_2 s_s k} \nn 
 & &   \cdot \frac{i e}{3} \, N_C \, N_{SF}
\eqn
with 
\bqn
{\cal O}_{\tilde m_{1_i} \tilde m_{2_i} m_s \tilde m_k}^{i; s_1
   s_2 s_s k} & = & 4 \pi \;\; \frac{\hat k_i}{\hat a_i} \,
                    (-1)^{l_{1_i}} \nn
   & & \cdot \hat a_i \, (-1)^{1-s_s+\half} \,
       \sdj{\half}{\half}{a_i}{s_s}{-s_2}{-m_s} \nn
   & & \cdot \hat{\tilde k}_i \, (-1)^{a_i-\tilde l_{1_i}+\tilde m_k} \,
       \sdj{a_i}{\tilde l_{1_i}}{\tilde k_i}{m_s}{\tilde m_{1_i}}
           {-\tilde m_k} \nn 
   & & \cdot \sqrt{3} \, (-1)^{\tilde l_{2_i}-\tilde k_i + k} \,
       \sdj{\tilde l_{2_i}}{\tilde k_i}{1}{\tilde m_{2_i}}{\tilde
         m_k}{-k} \nn
   & & \cdot \sqrt{2} \, (-1)^{S_1-\half+s} \, 
        \sdj{S_1}{\half}{\half}{s_1}{s_2}{-s} \nn
   & & \cdot \sqrt{2} \, (-1)^{S_1-\half+s'} \,
        \sdj{S_1}{\half}{\half}{s_1}{s_s}{-s'} 
\eqn
and the integral over the momentum p is (with the angular
dependence of $p_s^0, p_{\ov s}^0$ neglected) 
\bqn
{\cal I}_i & = & \frac{1}{(2 \pi \delta)^3} \; \int dp \, p^{2+t_i} \,
           \exp(-\alpha^2 \, p^2) \, 
           \left( - \frac{\sqrt{p_s^0+m_s}\sqrt{p_{\ov
                 s}^0+m_s}}{\sqrt{2 p_s^0 \, 2 p_{\ov s}^0}} \right)
           \nn
 & &       \hspace{2cm} \cdot \left\lbrace 
           \frac{1}{(p_s^0+m_s)(p_{\ov s}^0+m_s)} \right\rbrace_{i>1} 
           \quad.   
\eqn 
The spin-flavour and colour coefficients are 
(see appendix D) 
\bqn
N_{SF} & = & \sqrt{\frac{3}{2}} \quad, \\
N_C & = & \frac{1}{\sqrt{3}} \quad,
\eqn
and the coefficients $c_i$ are constant factors. \\
The averaged differential scattering cross section in the 
c.m.-frame is as usual
\bqn
\frac{d \sigma}{d \Omega} & = & \frac {1}{64 \, \pi^2 \, E_{CM}^2} \;
\frac{p_f}{p_i} \; \ov{|{\cal T}|^2} \nn
\mbox{ with } p_i = k = P_N & = & \half(E_{CM}-\frac{M_N^2}{E_{CM}}) \nn
             p_f = P_K = P_Y & = & \frac{1}{2 E_{CM}} \,
             ((E_{CM}^2-M_K^2-M_Y^2)^2 - 4 M_K^2 M_Y^2)^\half \quad.
\eqn

\section{Results}

\begin{figure}
\begin{center} 
\leavevmode
\epsfxsize=0.8\textwidth
\epsffile{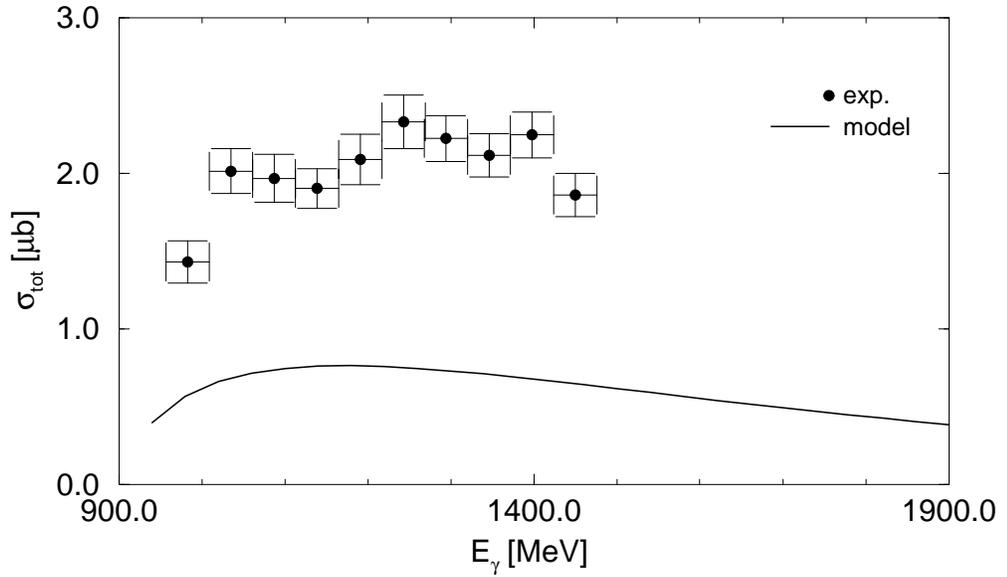}
\end{center}
\caption{The total cross section of the $\Lambda$ production;
  experimental data are taken from \protect\cite{schwille}}
\label{KLstot}
\end{figure}

\begin{figure}
\begin{center}
\leavevmode
\epsfxsize=0.7\textwidth
\epsffile{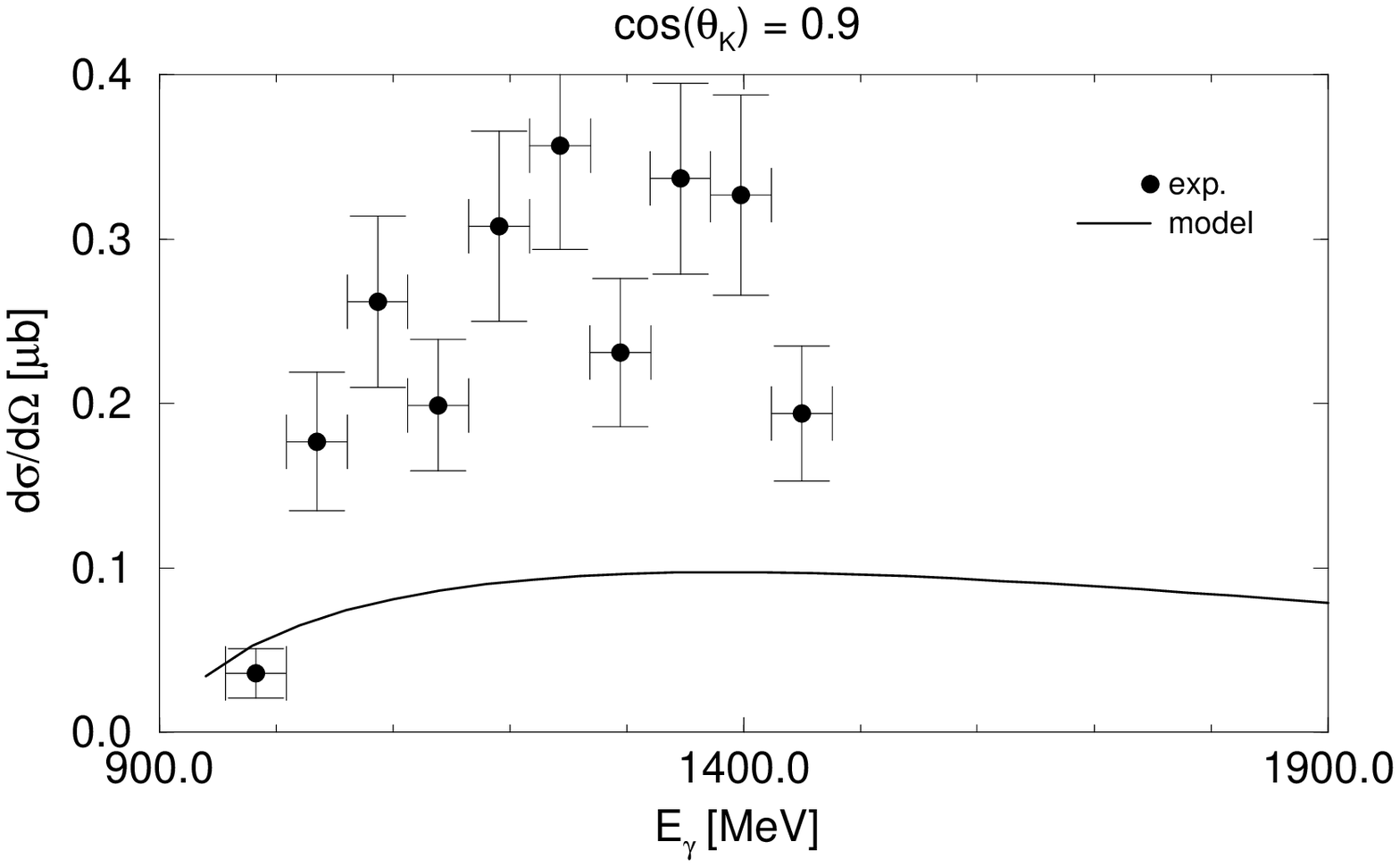}
\end{center}
\begin{center}
\leavevmode
\epsfxsize=0.7\textwidth
\epsffile{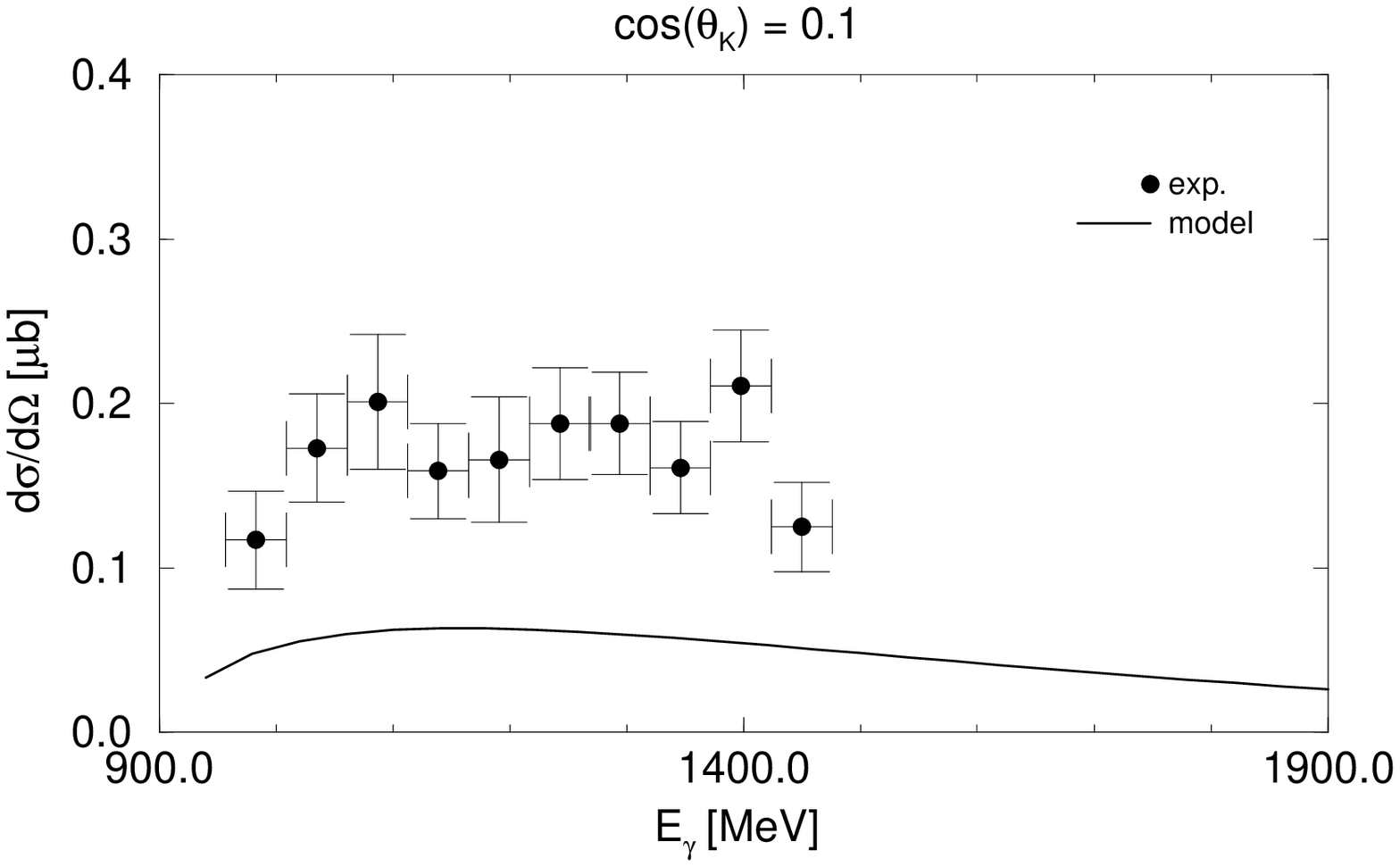}
\end{center}
\begin{center} 
\leavevmode
\epsfxsize=0.7\textwidth
\epsffile{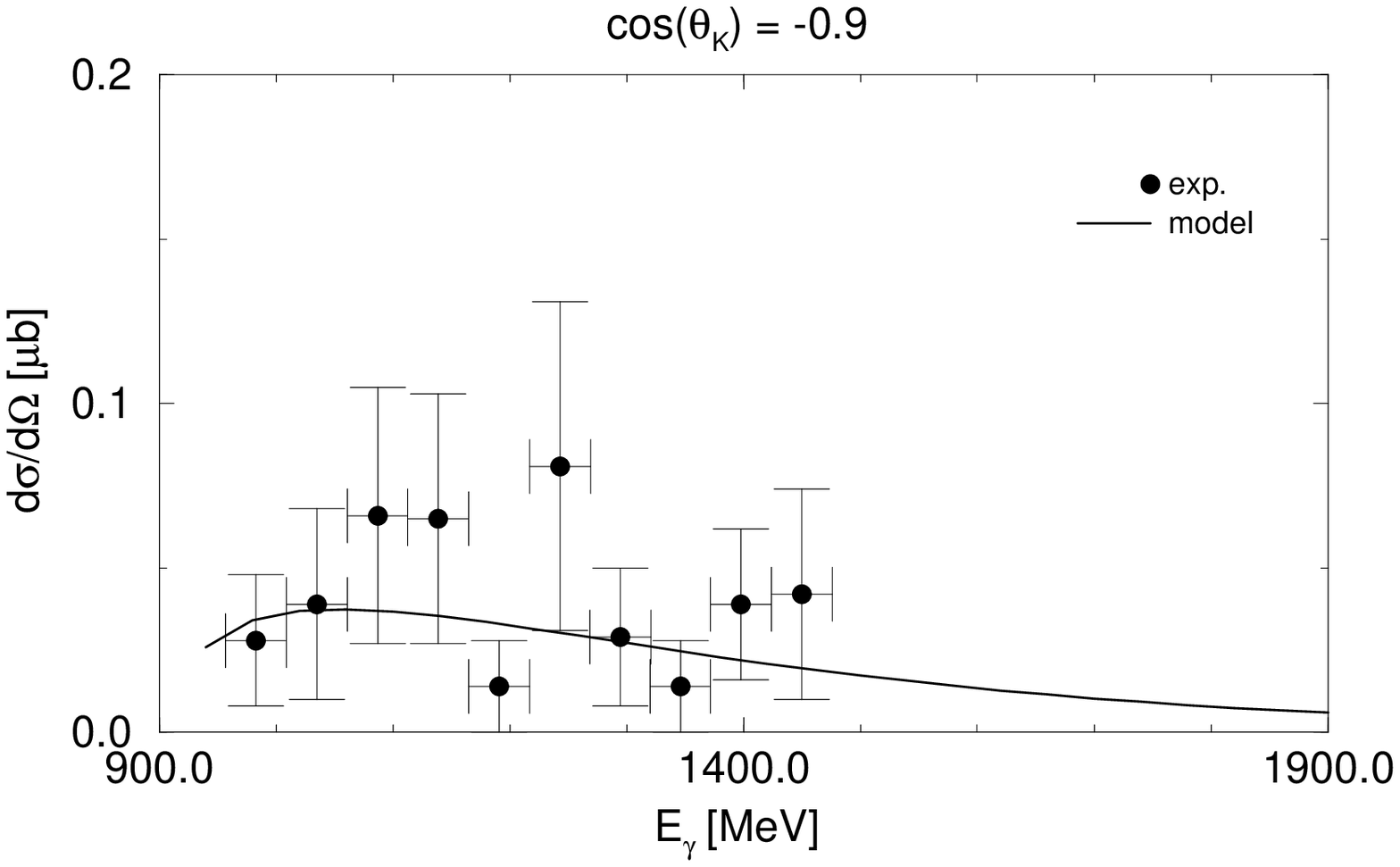}
\end{center}
\caption{The differential cross section of the $\Lambda$ production
  for three different scattering angles;
  experimental data are taken from \protect\cite{schwille}}
\label{KLcos}
\end{figure}

\begin{figure}
\begin{center} 
\leavevmode
\epsfxsize=0.7\textwidth
\epsffile{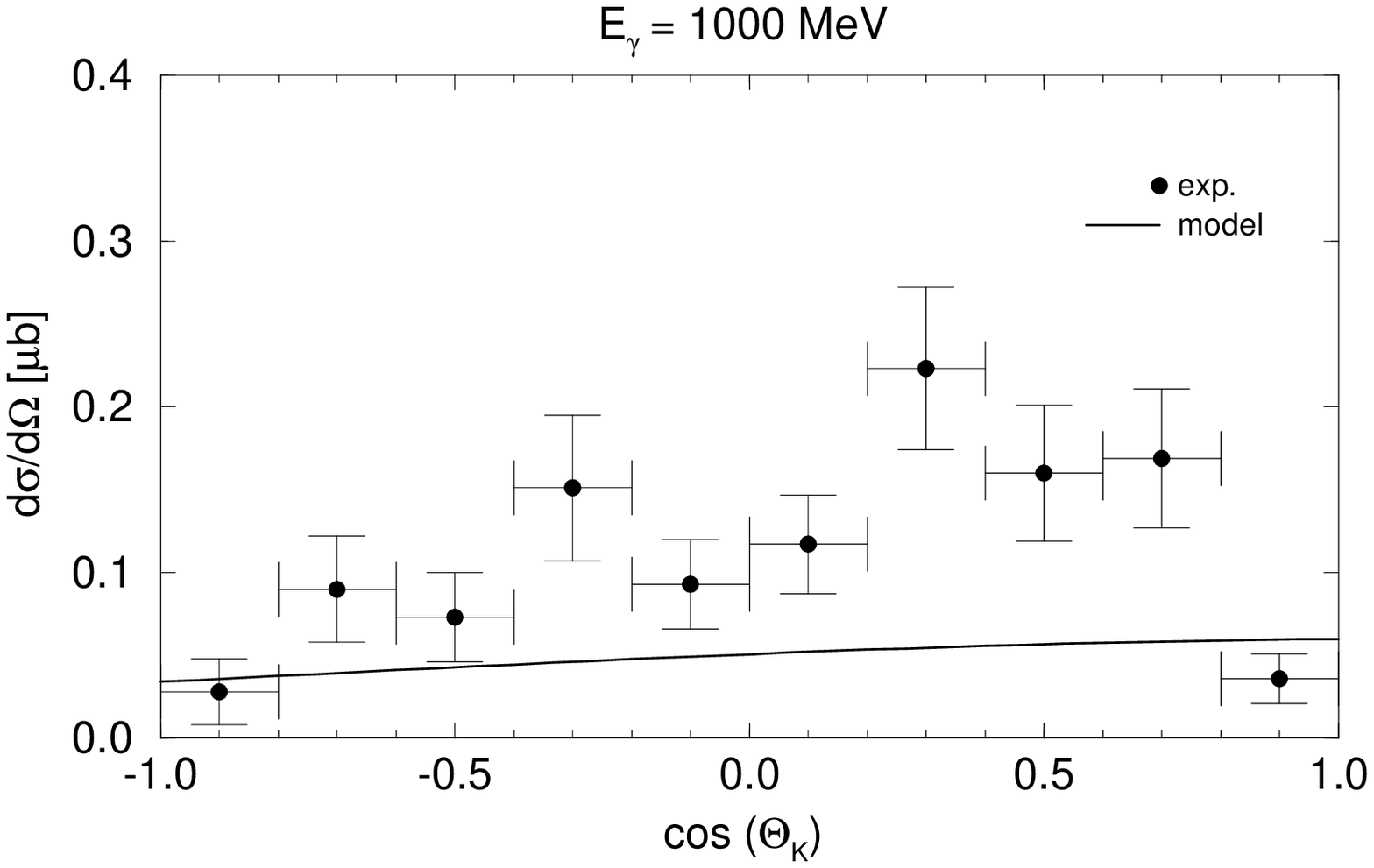}
\end{center}
\begin{center} 
\leavevmode
\epsfxsize=0.7\textwidth
\epsffile{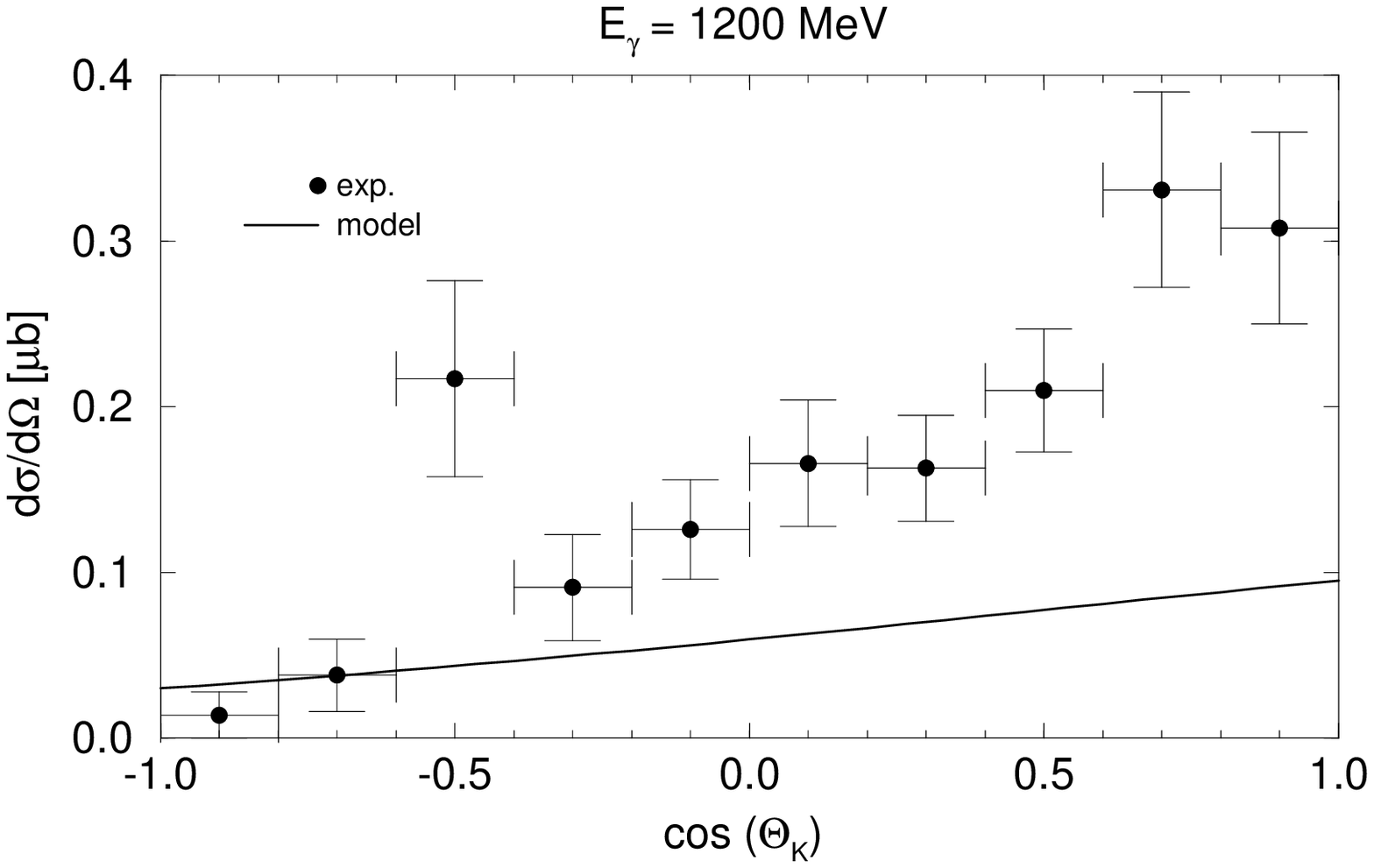}
\end{center}
\begin{center} 
\leavevmode
\epsfxsize=0.7\textwidth
\epsffile{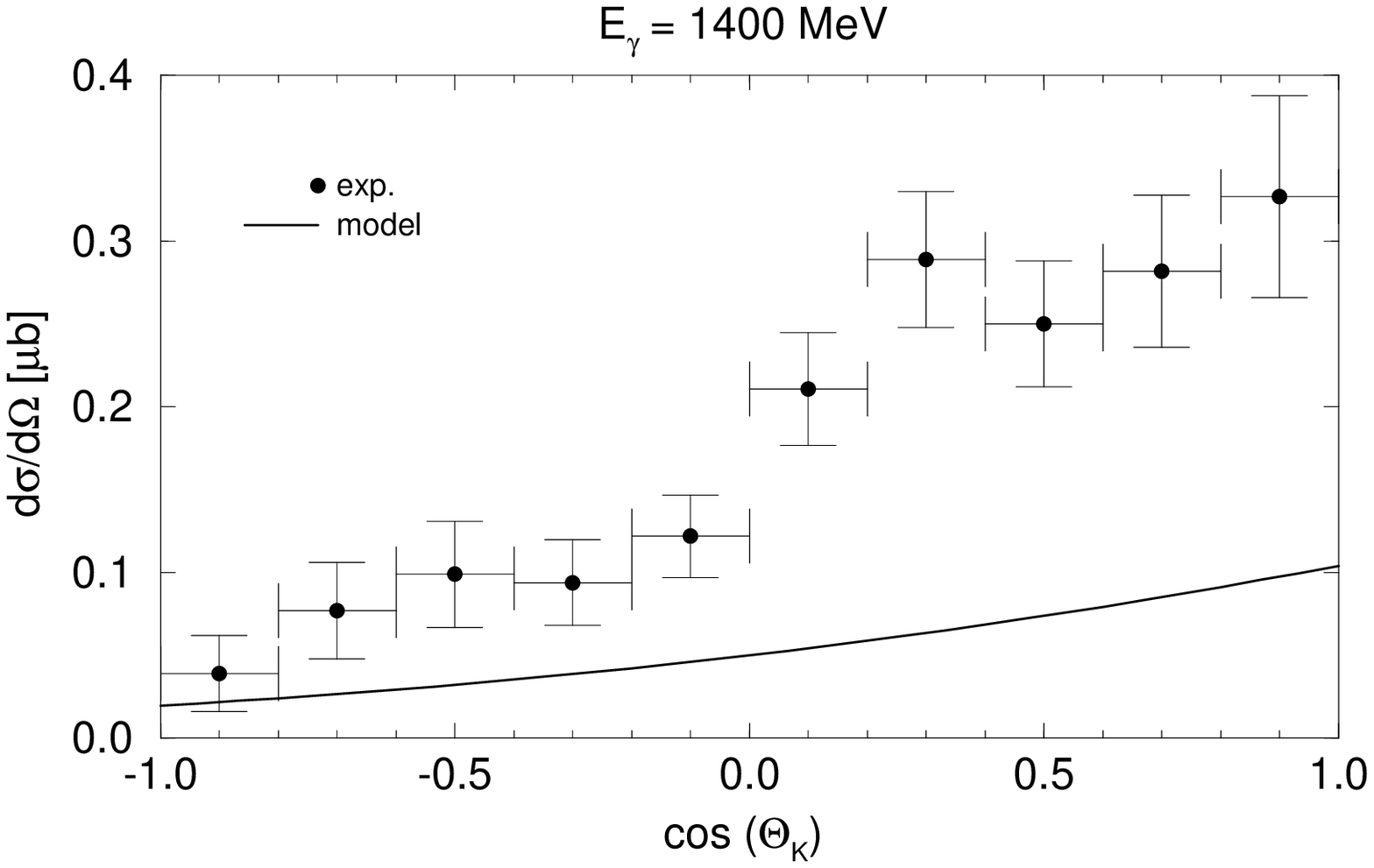}
\end{center}
\caption{The differential cross section of the $\Lambda$ production
  for three different photon energies;
  experimental data are taken from \protect\cite{schwille}}
\label{KLegam}
\end{figure}

\begin{figure}
\begin{center} 
\leavevmode
\epsfxsize=0.7\textwidth
\epsffile{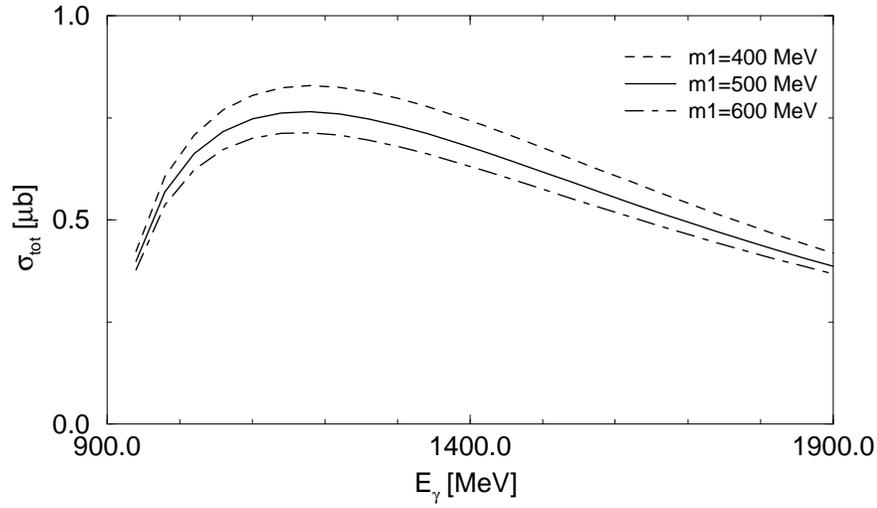}
\end{center}
\caption{The total cross section of the $\Lambda$ production for three
  different values of the spectator diquark mass $m_1$}
\label{KLvarm1}
\end{figure}

\begin{figure}
\begin{center} 
\leavevmode
\epsfxsize=0.7\textwidth
\epsffile{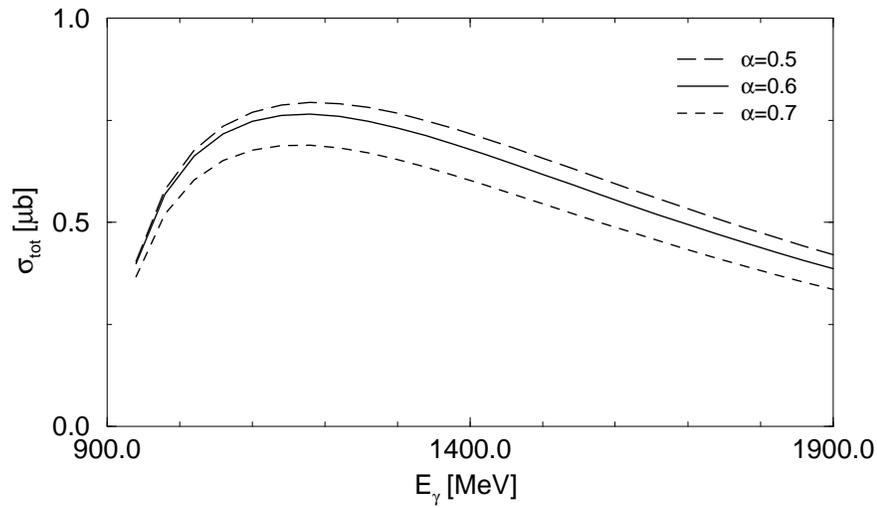}
\end{center}
\caption{The total cross section of the $\Lambda$ production for three
  different values of the nucleon oscillator parameter $\alpha$}
\label{KLvaralpha}
\end{figure}

For the following values of the parameters we find a
semi-quantitative description of the differential and total 
scattering cross sections:
\bqn 
m_n & = & 300 \; \mbox{MeV} \quad,\quad m_s = 500 \; \mbox{MeV} \quad,\quad
m_1(\Lambda) = 500 \; \mbox{MeV} \quad,\quad \nn
\alpha & = & 0.6 \; \mbox{fm} \quad,\quad 
\beta = 0.4  \; \mbox{fm} \quad,\quad 
\gamma = 0.3 \; \mbox{fm}
\quad .  
\eqn   
The values of the constituent quark masses are in a range
adopted in most quark models (see e.g. \cite{weise}).
The spectator diquark mass is chosen a little less than the sum of 
u and d quarks \cite{lichtenberg}. The oscillator parameters
decribing the extension of the hadrons are in a physical
acceptable region (see also \cite{weise} for models predicting
the nucleon and pion quark-core radii). The radii of the
strange hadrons may follow from the relation motivated 
by an oscillator potential: $\frac{r_2}{r_1} = \sqrt{\frac{M_1}{M_2}}$. \\ 
Fig.\ref{KLstot} to \ref{KLegam} show the results of our
model in comparison to experimental data taken from
\cite{schwille}. The broad maximum of the
total cross section between
$E_\gamma= 1.0 \; \mbox{GeV}$ and $1.4 \; \mbox{GeV}$ 
as well as the probable decrease for energies greater than $1.4 \;
\mbox{GeV}$ \cite{schoch} is qualitatively 
reproduced very well. However, the calculated curve is too 
low by a factor of about $2.5$. In fact, a variation of the parameters
cannot increase the calculated results. Of course a correct
quantitative decription of the experimental data was not 
expected. This is due to the simplicity of the model; 
we simply neglect the coupling of the photon to the 
spectator quarks. 
Although the calculated differential cross sections are too small,
the backward scattering ($\cos \, (\theta_K)<1$) is
described almost quantitatively. The rise of the differential 
cross section
with increasing $\cos \, (\theta_K)$ is reproduced, too. \\
Fig.\ref{KLvarm1} shows the calculated total cross section
for different values of the spectator diquark mass $m_1$.
Fig.\ref{KLvaralpha} shows the model prediction for different
proton oscillator parameters.
Apparently, the variation of the results is modest, once 
the parameters have been chosen in a physical acceptable range. It is
interesting to see that a smaller diquark mass has almost the same
effect as a smaller nucleon oscillator parameter, as is expected
for a larger binding energy. 

\section{Summary and Outlook}
We developed a simple quark model to describe qualitatively and
semi-quantitatively the differential and total cross sections 
of the reaction 
$\gamma + p \rightarrow K^+ + \Lambda$ for photon energies up to
$1.9 \; \mbox{GeV}$.
The only contribution
considered is the coupling of the photon to the
strange quark and antiquark in the final state, which recombine
with the spectator quarks of the proton. Apparently, 
this process has a large contribution to the cross section. 
It would be interesting to study the contribution of the
spectator quarks to the reaction, which cannot be 
calculated in our model. A main drawback of the model is
that it is not able to calculate the recoil
polarization, which is zero since the following
relation holds for the ${\cal T}$ matrix:
\bqn
{\cal T}_{s s' \lambda} = (-1)^{s+s'+\lambda} \; {\cal T}_{-s -s'
  -\lambda}^{\;\;*} \quad.
\eqn
To obtain a non-vanishing polarization one may include 
hadronic resonances.

In a future calculation, the photoproduction of both 
$\Lambda$ and $\Sigma^0$ will be examined in a fully 
relativistic model. The transition matrix will be calculated
in the Mandelstam \mbox{formalism} \cite{mandelstam}, and the
hadrons are described as relativistically bound states of quarks
through Bethe-Salpeter amplitudes. In this framework, all 
three diagrams, where the photon couples to all possible internal 
quark lines, are taken into account. This gives rise to
interferences. 
In addition, the amplitudes
are properly boosted, which cannot be neglected in 
kaon-photoproduction.

{\bf Acknowledgements:} I am grateful to H.R. Petry, B.C. Metsch,
C.R. M\"unz and J. Resag for many helpful discussions. This work was
supported by the Deutsche Forschungsgemeinschaft.

\newpage

\newpage

\newpage

\newpage

\begin{appendix}

\section{Normalization of the bound states}
\label{hadnorm}

\noindent
The states
\bqn
\ket{N_{s}(P_N)} = N_\alpha \sqrt{2 P_N^0} \; \intp{p} \, 
 \frac{1}{\sqrt{2p_1^0 \,  2p_2^0}} \, 
 {\cal R}_\alpha(p) \chi_N^F \chi_N^C \, \ket{\vec p_1} \ket{\vec p_2, s}
\eqn
are normalized according to:
\bqn
\braket{N_{s}(P_N)}{N_{s'}(P'_N)} = (2 \pi)^3 \, (2 P_N^0) 
                          \delta^3(\vec P_N - \vec P'_N) \, 
                          \delta_{s s'} \; \delta_{F F'} \quad.
\eqn
It follows:
\bqn
N_\alpha = 4 \pi \, \alpha^{\frac{3}{2}} \,
\left(\frac{2}{\pi}\right)^\frac{1}{4} \quad. 
\eqn
Analogous for $N_\beta$ and $N_\gamma$.

\section{Decomposition of field operators, commutator relations}
\label{fieldop}

\noindent
The quark and photon field operators are
\bqn
\Psi(0) & = & \intp{\tilde{\tilde p}} \frac{1}{2 \tilde{\tilde p}^0} 
   \sum_{\tilde{\tilde s}} \left( 
   v_{\tilde{\tilde s}}(\tilde{\tilde p}) \, b_{\tilde{\tilde
     s}}^{\dag}(\tilde{\tilde p}) + u_{\tilde{\tilde s}} (\tilde{\tilde p})
   \, \tilde{\tilde a}_{\tilde{\tilde s}}(\tilde{\tilde p}) \right) \nn
\ov \Psi(0) & = & \intp{\tilde p} \frac{1}{2 \tilde p^0} 
   \sum_{\tilde s} \left( 
   \ov v_{\tilde s}(\tilde p) \, b_{\tilde s}(\tilde p) + \ov u_{\tilde
     s} (\tilde p)
   \, \tilde a_{\tilde s}^{\dag}(\tilde p) \right) \nn
\lbrace b_{\tilde s}(\tilde p) , b_{\tilde{\tilde
    s}}^{\dag}(\tilde{\tilde p}) \rbrace & = &
\lbrace a_{\tilde s}(\tilde p) , a_{\tilde{\tilde
    s}}^{\dag}(\tilde{\tilde p}) \rbrace =  
(2 \pi)^3 \, (2 {\tilde p}^0) \, \delta^3(\tilde{\vec
  p}-\tilde{\tilde{\vec p}}) \, \delta_{\tilde s \tilde{\tilde s}} \nn
A_\mu(0) & = & \intp{k} \, \frac{1}{2 k^0} \, \sum_{\lambda=1}^2 \,
\eps_\mu^{(\lambda)}(k) \, (a^{(\lambda)}(k) +
{a^{(\lambda)}}^{\dag}(k)) \nn
\rightarrow \; A_\mu(0) \, \ket{(\vec k, \lambda)} & = & 
\eps_\mu^{(\lambda)} (k) \, \ket{0} \nn 
{[ a^{(\lambda)}(k) , {a^{(\lambda')}}^{\dag}(k') ]} & = & 
(2 \pi)^3 \,  
(2 k^0) \, \delta^3(\vec k - \vec k \,') \, \delta_{\lambda \lambda'}
\quad,\quad \lambda=1\;, 2 
\eqn
and the bosonic operators fulfill:
\bqn   
{[ \tilde a_0(p), \tilde a_0^{\dag}(p') ]} & = & 
(2 \pi)^3 \,  
(2 p^0) \, \delta^3(\vec p - \vec p \,')  \quad.   
\eqn

\section{Coordinates}

\noindent
The (di-)quark coordinates are
\bqn
\vec p_1 & = & \tilde m_1 \, \vec P_N + \vec p \nn
\vec p_2 & = & \tilde m_n \, \vec P_N - \vec p \nn
\vec p_1\,' & = & \tilde{\tilde m}_1 \, \vec P_Y +\vec p\,' \nn
\vec p_s & = & \frac{m_s}{m_1+m_s} \, \vec P_Y -\vec p\,' \nn
\vec p_{\ov s} & = & \frac{m_s}{m_s+m_n} \, \vec P_K +\vec p\,'' \nn
\vec p_2\,' & = & \tilde{\tilde m}_n \, \vec P_K -\vec p\,''  
\quad.
\eqn
From
$ \vec p_1  =  \vec p_1\,'$ and  
$\vec p_2  =  \vec p_2\,' $
follows
\bqn
\vec p\,' & = & \tilde m_1 \vec P_N - \tilde{\tilde m}_1 \vec P_Y +
\vec p \nn 
\vec p\,'' & = & \tilde{\tilde m}_n \vec P_K - \tilde m_n \vec P_N +
\vec p \quad. 
\eqn

\section{Symmetry Coefficients} \label{symfac}

\noindent
The three-quark wave functions of the baryons are
\bqn
\Psi(1,2,3) = \psi_{space}^S (1,2,3) \; (\chi_{spin} \otimes
\phi_{flavour})^S (1,2,3)
\; \chi_{colour}^A (1,2,3) \quad.
\eqn
The spin-flavour function is symmetric under the
interchange of any two quarks:
\bqn
\ket{p} & = & \frac{1}{\sqrt{2}} \; \left( \; \chi_{M_S} \;
\lbrack\lbrack n \; n\rbrack^1 n \rbrack^\half + \chi_{M_A} \; 
\lbrack\lbrack n \; n\rbrack^0 n \rbrack^\half \; \right) \nn
\ket{\Lambda} & = & \half \; \chi_{M_S} \; \left( \;  
    \lbrack \lbrack s n\rbrack^\half \; n\rbrack^0 \; + \;
    \lbrack \lbrack n s\rbrack^\half \; n\rbrack^0 \; \right) \nn
& & + \chi_{M_A} \; 
\left(-\frac{1}{\sqrt{12}} \; 
\lbrack \lbrack s n\rbrack^\half \; n\rbrack^0 \; +
  \frac{1}{\sqrt{3}} \;  
\lbrack \lbrack n n\rbrack^0 \; s\rbrack^0 \; +
  \frac{1}{\sqrt{12}} \;
\lbrack \lbrack n s\rbrack^\half \; n\rbrack^0 \; \right) \nn
\ket{\Sigma} & = & -\frac{1}{\sqrt{12}} 
\chi_{M_S}\lbrack\lbrack n s \rbrack^\half n \rbrack^1 +
\frac{1}{\sqrt{3}} \chi_{M_S}\lbrack\lbrack n n \rbrack^1 s \rbrack^1 
-\frac{1}{\sqrt{12}} 
\chi_{M_S}\lbrack\lbrack s n \rbrack^\half n \rbrack^1 \nn
& & + \half \chi_{M_A}\lbrack\lbrack n s \rbrack^\half n \rbrack^1 
- \half \chi_{M_A}\lbrack\lbrack s n \rbrack^\half n \rbrack^1 
\eqn 
with e.g.
\bqn
\lbrack\lbrack n n \rbrack^0 n \rbrack^\half & = & \frac{1}{\sqrt{2}}
(ud-du)u \nn 
\lbrack\lbrack n n \rbrack^1 n \rbrack^\half & = & \frac{1}{\sqrt{6}}
((ud+du)u - 2uud) \nn
\lbrack\lbrack n n \rbrack^1 s \rbrack^1 & = & \frac{1}{\sqrt{2}}
(ud+du)s   \quad.
\eqn 
The $\chi_{M_A}, \chi_{M_S}$ are the (in quarks 1-2) mixed
(anti-)symmetric spin functions:
\bqn
\chi_{M_A} & = & \left\lbrack\left\lbrack \half \otimes \half
\right\rbrack^0 \otimes\half\right\rbrack^S  \nn 
\chi_{M_S} & = & \left\lbrack\left\lbrack \half \otimes \half
\right\rbrack^1 \otimes\half\right\rbrack^S   
\quad.
\eqn
For the $\Lambda$ production one gets from
\bqn
3 \; \braket{\frac{1}{\sqrt{3}} \; \chi_{M_A}\; 
\lbrack \lbrack n n \rbrack^0 s \rbrack^0 }
            {\frac{1}{\sqrt{2}} \; \chi_{M_A}\;
\lbrack \lbrack n n \rbrack^0 n \rbrack^\half }
\eqn
the factor $N_{SF} = \sqrt{\frac{3}{2}}$, 
for the $\Sigma$ production from
\bqn
3 \; \braket{\frac{1}{\sqrt{3}} \; \chi_{M_S}\; 
\lbrack \lbrack n n \rbrack^1 s \rbrack^1 }
            {\frac{1}{\sqrt{2}} \; \chi_{M_S}\;
\lbrack \lbrack n n \rbrack^1 n \rbrack^\half }
\eqn
the same factor $N_{SF} = \sqrt{\frac{3}{2}}$.
The colour factor results from the colour singlet functions
of the hadrons
\bqn
\ket{\mbox{Baryon}}_{col} & = & \frac{1}{\sqrt{6}} (rgb-rbg+gbr-grb+brg-bgr)
\nn
\ket{\mbox{Meson}}_{col} & = & \frac{1}{\sqrt{3}} (\ov r r + \ov g g + \ov b
b)
\eqn
to
\bqn
N_{C} = 3 \; \frac{1}{\sqrt{3}} \; \frac{1}{\sqrt{6}} \; 2 \;
\frac{1}{\sqrt{6}} = \frac{1}{\sqrt{3}} \quad.
\eqn

\end{appendix}

\end{document}